\documentclass{article}
\usepackage{amsfonts,amssymb,amsmath,graphicx,hyperref}
\begin{document}

\title{New Mechanics of Generic Musculo-Skeletal Injury}
\author{Vladimir G. Ivancevic \\
Defence Science \& Technology Organisation, Australia}
\date{}
\maketitle

\begin{abstract}
Prediction and prevention of musculo-skeletal injuries is an important
aspect of preventive health science. Using as an example a human knee joint, this paper proposes a new
\emph{coupled--loading--rate hypothesis}, which states that a
generic cause of any musculo-skeletal injury is a \textsl{Euclidean jolt}, or
$SE(3)-$jolt, an impulsive loading that hits a joint in several
coupled degrees-of-freedom simultaneously. Informally, it is a
rate-of-change of joint acceleration in all 6-degrees-of-freedom simultaneously, times the corresponding portion of the body mass. In the case of a human knee, this happens when most of the body mass is on one leg
with a semi-flexed knee -- and then, caused by some external
shock, the knee suddenly `jerks'; this can happen in running,
skiing, sports games (e.g., soccer, rugby) and various
crashes/impacts. To show this formally, based on the previously
defined \emph{covariant force law} and its application to traumatic brain injury \cite{GaneshTBI}, we formulate the coupled
Newton--Euler dynamics of human joint motions and derive from it the
corresponding coupled $SE(3)-$jolt dynamics of the joint in case. The $SE(3)-$jolt is
the main cause of two forms of discontinuous joint injury: (i) mild
rotational disclinations and (ii) severe translational
dislocations. Both the joint {disclinations and dislocations}, as
caused by the $SE(3)-$jolt, are described using the Cosserat
multipolar viscoelastic continuum joint model. \bigbreak

\end{abstract}

\vspace{5cm}

\noindent\textbf{Contact information:}\bigbreak

\noindent Dr. Vladimir Ivancevic, Senior Research Scientist,\newline
Human Systems Integration, Land Operations Division\newline
Defence Science \& Technology Organisation, Australia\newline
PO Box 1500, 75 Labs, Edinburgh SA 5111\newline
Tel:~ +61 8 8259 7337, ~~~Fax:~ +61 8 8259 4193\newline
E-mail: ~ Vladimir.Ivancevic$@$dsto.defence.gov.au


\newpage

\section{Introduction}

\begin{figure}[htb]
\centerline{\includegraphics[width=8cm]{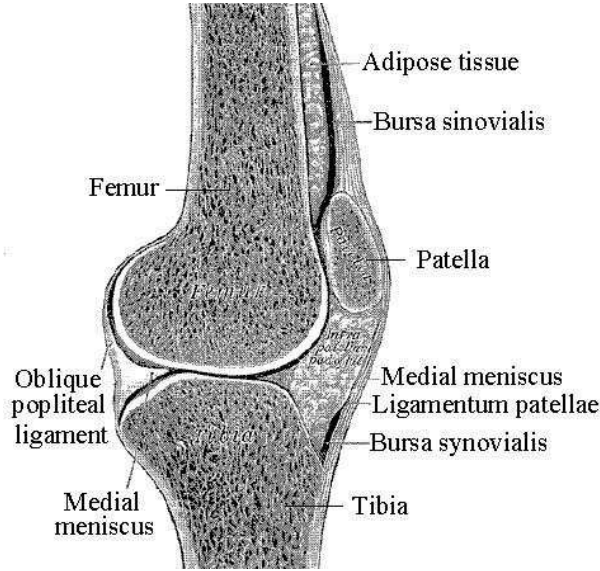}}\caption{ Schematic
latero-frontal view of the left knee joint. Although designed to
perform mainly flexion/extension (strictly in the sagittal plane)
with some restricted medial/lateral rotation in the semi-flexed
position, it is clear that the knee joint really has \emph{at
least} 6 degrees-of-freedom, including 3 micro--translations. The
injury actually occurs when some of these \emph{microscopic}
translations become \emph{macroscopic}, which normally happens
only after an external jolt.}\label{knee}
\end{figure}

In this paper, we propose a new model of musculo-skeletal injury, using as an example the human knee joint. The knee joint comprises three articulations: (i) a tibio-femoral
joint) between the medial and lateral condyles of the femur and
tibia (see Figure \ref{knee}), (ii) patelo-femoral joint between
the femur and patella, and (iii) tibio-fibular joint between tibia
and fibula. The knee is double condyloid joint, with a dominant
flexion/extension. As a synovial joint, the knee has strong
fibrous capsule that attaches superiorly to the femur and
inferiorly to the articular margin of the tibia. Given its
relatively poor bony fit, the knee relies on ligaments for much of
its structural stability and integrity. \cite{Whiting}.

Knee joint injuries range from mild ligament or meniscus tearing
to severe traumatic dislocations (see \cite{Seroyer} and
references therein) that fall among the most severe form of
ligament injury to the lower extremity, associated with a high
rate of complications including amputation.

Knee joint injuries are also frequent in sports, especially in
ball games. For example, in a recently case--reported complex knee
injury in a rugby league player \cite{Shillington}, resulting in
combined rupture of the patellar tendon, anterior cruciate and
medial collateral ligaments, with a medial meniscal tear, the
video--analysis suggests two points during the tackle when there
was the potential for injury: the first occurred when the player
was in single leg stance whilst running and received impact to his
upper body from three defenders; the second occurred when the
player landed on the knee and sustained a valgus and hyper-flexion
force under the weight of two defenders. The goal of treatment in
this condition is restoration of both the extensor mechanism and
knee stability.

Also, the increased number of women participating in sports like
soccer has been paralleled by a greater knee injury rate in women
compared to men. In particular, menstrual cycle phase has been correlated with risk of noncontact anterior cruciate ligament injury in women (see, e.g. \cite{Chaudhari}).\footnote{According to \cite{Chaudhari}, jumping and landing activities performed during different phases of the menstrual cycle lead to differences in foot strike knee flexion, as well as peak knee and hip loads, in women not taking an oral contraceptive but not in women taking an oral contraceptive. Women will experience greater normalized joint loads than men during these activities.} Among these injuries, those occurring to the
anterior cruciate ligament are commonly observed during sidestep
cutting maneuvers \cite{Sanna}. In addition, general fatigue
appears to correlate with injuries to the passive knee--joint
structures during a soccer game. It has been observed that a higher injury rate and more severe injuries occur towards the end of a soccer game or practice \cite{Hawkins,Ostenberg}, suggesting a fatigue effect on the neuromuscular system. Besides, women soccer players who sustain knee injuries have a high risk of going on to develop osteoarthritis at a young age \cite{Lohmander}.

Besides, the knee is the body part most commonly injured as a
consequence of collisions, falls, and overuse occurring from
childhood sports. The number of sports--related injuries is
increasing because of active participation of children in
competitive sports \cite{Siow}. Children differ from adults in many areas,
such as increased rate and ability of healing, higher strength of
ligaments compared with growth plates, and continued growth.
Growth around the knee can be affected if the growth plates are
involved in injuries \cite{Siow}.

Literature on knee mechanics has been reviewed in \cite{Komistek},
evaluating various techniques that had been used to determine in
`vivo loads' in the human knee joint. Two main techniques that had
been used were 3-axial accelerometer-based telemetry -- an experimental approach, and
mathematical modelling -- a theoretical approach. Accelerometric
analyzes had previously been used to determine the `in vivo'
loading of the human hip and more recently evaluated in the
determination of in vivo knee loads. Mathematical modelling
approaches can be categorized in two ways: (i) those that use
optimization techniques to solve an indeterminate system, and (ii)
those that utilize a reduction method that minimizes the number of
unknowns, keeping the system solvable as the number of equations
of motion are equal to the number of unknown quantities (for more
technical details, see \cite{Komistek} and references therein).

The use of a \emph{force--controlled dynamic knee simulator} to
quantify the mechanical performance of total knee replacement
designs during functional activity was pioneered by
\cite{DesJardins1}, in which dynamic \emph{total knee replacement}
(TKR) study utilized a 6-degree-of-freedom force--controlled knee
simulator to quantify the effect of TKR design alone on TKR
mechanics during a simulated walking cycle. Simultaneous
prediction of implant kinematics and contact mechanics has been
demonstrated using explicit \emph{finite element} (FE) models of
the \emph{Instron/Stanmore Knee Joint Simulator, Instron, Canton,
MA} \cite{Godest,Halloran,Halloran2} In these models, kinematic
verification was performed by comparing experimental and
model-predicted motion for a single implant. Both models were
found to produce similar kinematic simulation results. Estimated contact
pressure distributions were also closely correlated, as long as
significant edge--loading conditions were not present. Recently,
an adaptive FE method for pre--clinical wear testing of TKR
components was developed in \cite{Knight}, capable of simulating
wear of a polyethylene tibial insert and to compare predicted
kinematics, weight loss due to wear, and wear depth contours to
results from a force--controlled experimental knee simulator. The
displacement--controlled inputs, by accurately matching the
experimental tibio-femoral motion, provided an evaluation of the
simple wear theory. The force--controlled inputs provided an
evaluation of the overall numerical method by simultaneously
predicting both kinematics and wear. Proposed international
standards for TKR wear simulation have been drafted (see \cite{ISO Standard}), yet their
methods continue to be debated \cite{Laz}. The `gold standard' to which all
TKR wear testing methodologies should be compared is measured in
vivo TKR performance in patients. The study of \cite{DesJardins}
compared patient TKR kinematics from fluoroscopic analysis and
simulator TKR kinematics from force--controlled wear testing to
quantify similarities in clinical ranges of motion and contact
bearing kinematics and to evaluate the proposed ISO
force--controlled Stanmore wear testing methodology.

For human movement purposes, we can say that the safe knee motions
(flexion/extension with some medial/lateral rotation in the flexed
position) \textit{are} governed by standard Euler's rotational
dynamics coupled to Newton's micro-translational dynamics. On the
other hand, the unsafe knee events, in this paper we will show that the main cause of knee
injuries are the knee SE(3)--jolts, the sharp and
sudden, \textquotedblleft delta\textquotedblright -- (forces +
torques) combined. These knee SE(3)--jolts do not belong to the
standard Newton--Euler dynamics. The only way to monitor them
would be to measure \textquotedblleft in vivo" the rate of the
combined (forces + torques)-- rise in the knee joint (see Figure
\ref{knee}).\\

This paper proposes a new hypothesis for generic musculo-skeletal injury, called:\\

\noindent {\bf Coupled--loading--rate hypothesis:}\\
\emph{The main cause of knee injury is {\bf Euclidean
jolt}, or $SE(3)-$jolt, an impulsive loading that hits the knee
joint in several coupled degrees-of-freedom (DOF) simultaneously. The same hypothesis applies to any other major human joint: all acute musculo-skeletal injuries are caused by some form of Euclidean jolt.} \\

In the realm of musculo-skeletal injury, a Euclidean
jolt represents a 6-degree-of-freedom `jerk' (rate-of-change of
acceleration) times most-of-the-body mass. In other words, a Euclidean
jolt is a time derivative of the Euclidean force (three-dimensional force + three-dimensional torque). In the case of a knee, it happens when all
the body mass is on one leg with a semi-flexed knee -- and then,
caused by some external shock, the knee `jerks'; this often happens in running,
skiing, sports games (e.g., soccer, rugby) and various
crashes/impacts. The reason why this happens in this particular scenario is the following: in a semi-flexed position, the knee joint has all 6 DOF;\footnote{The frequently used human leg model by \cite{Brand} was comprised of 47 muscles and each joint was represented by three interactive forces and three interactive torques \cite{Komistek} -- thus clearly showing a generic 6 DOF function. In the case of the knee joint, the macroscopic movement of patella during flexion/extension clearly shows the existence of micro-translations in this joint (see also explanation in the caption of Figure 1). As this paper tries to show generic musculo-skeletal injury patterns, note that similar 6 DOF function is even more prominent in the case of human shoulder.} if a sudden jerk with the full body mass hits the knee in this position, it will happen in all 6 DOF simultaneously, tearing apparat the soft tissue. If the resulting jolt is of high intensity then even the hard tissue will be damaged.

To demonstrate this formally, based on the previously defined \emph{covariant force
law}, we formulate the coupled Newton--Euler dynamics of the knee
motions and derive from it the corresponding coupled $SE(3)-$jolt
dynamics. The effect of $SE(3)-$jolt can be seen in two forms of
discontinuous knee injury: (i) mild rotational disclinations and
(ii) severe translational dislocations. Both the knee
{disclinations and dislocations}, as caused by the $SE(3)-$jolt,
are described using the Cosserat multipolar viscoelastic continuum
model.

While we can intuitively visualize the knee SE(3)--jolt, for the purpose of whole human musculo-skeletal dynamics simulation, to avoid deterministic chaos caused by nonlinear coupling, we use the necessary simplified, decoupled approach (neglecting
the 3D torque matrix and its coupling to the 3D force vector). In this decoupled framework of reduced complexity, we define:

The cause of knee dislocations is a linear 3D--jolt vector, the time
rate-of-change of a 3D--force vector (linear jolt = mass $\times $ linear
jerk). The cause of knee disclinations is an angular 3--axial jolt, the time
rate-of-change of a 3--axial torque (angular jolt = inertia moment $\times $
angular jerk).

This decoupled framework has been implemented in the Human Biodynamics
Engine \cite{18}, a high-resolution neuro--musculo--skeletal dynamics simulator
(with 270 DOFs, the same number of equivalent muscular actuators and
two--level neural reflex control), developed by the present author at
Defence Science and Technology Organization, Australia. This kinematically
validated human motion simulator has been described in a series of papers
and books \cite{6,7,8,9,11},\newline
\cite{12,13,14},\newline
\cite{14a,17,IJHR,NeuFuz,StrAttr,CompMind,Complexity}.

\section{The $SE(3)-$jolt: the main cause of human joint injury}

In the language of modern biodynamics \cite{9,12}, the general knee motion is
governed by the Euclidean SE(3)--group of 3D motions. Within the
knee SE(3)--group we have both SE(3)--kinematics (consisting of
the knee SE(3)--velocity and its two time derivatives:
SE(3)--acceleration and SE(3)--jerk) and the knee SE(3)--dynamics
(consisting of SE(3)--momentum and its two time derivatives:
SE(3)--force and SE(3)--jolt), which is the knee kinematics
$\times $ the knee mass--inertia distribution.

Informally, the \textit{knee SE(3)--jolt}\footnote{%
The mechanical SE(3)--jolt concept is based on the mathematical
concept of higher--order tangency (rigorously defined in terms of
jet bundles of the head's configuration manifold) \cite{14,17}, as
follows: When something hits the human head, or the head hits some
external body, we have a collision. This is naturally described by
the SE(3)--momentum, which is a nonlinear coupling of 3 linear
Newtonian momenta with 3 angular Eulerian momenta. The tangent to
the SE(3)--momentum, defined by the (absolute) time derivative, is
the SE(3)--force. The second-order tangency is given by the
SE(3)--jolt, which is the tangent to the SE(3)--force, also
defined by the time derivative.} is a sharp and sudden change in
the SE(3)--force acting on the mass--inertia distribution within
the knee joint. That is, a `delta'--change in a 3D force--vector
coupled to a 3D torque--vector, hitting the knee. In other words,
the knee SE(3)--jolt is a sudden, sharp and discontinues shock in
all 6 coupled dimensions of the knee joint, within the three
Cartesian ($x,y,z$)--translations and the three corresponding
Euler angles around the Cartesian axes: roll, pitch and yaw
\cite{7}. If the SE(3)--jolt produces a mild shock to the knee (internal sudden loss of stability), it causes mild, soft--tissue knee injury. If the SE(3)--jolt produces
a hard shock (external hit by a massive body) to the knee, it causes severe, hard--tissue knee injury, with the total loss of knee movement.

The knee SE(3)--jolt is rigorously defined in terms of
differential geometry\newline \cite{14,17}. Briefly, it is the
absolute time--derivative of the covariant force 1--form (or,
co-vector field) applied to the knee. With this respect, recall
that the fundamental law of biomechanics -- the so--called
\emph{covariant force law} \cite{13,14,17}, states:
\begin{equation*}
\text{Force co-vector field}=\text{Mass distribution}\times \text{%
Acceleration vector--field},
\end{equation*}%
which is formally written (using the Einstein summation convention, with
indices labelling the three local Cartesian translations and the
corresponding three local Euler angles):
\begin{equation*}
F_{{\mu }}=m_{{\mu }{\nu }}a^{{\nu }},\qquad ({\mu ,\nu }=1,...,6=3\text{
Cartesian}+3\text{ Euler})
\end{equation*}%
where $F_{{\mu }}$ denotes the 6 covariant components of the knee
SE(3)--force co-vector field, $m_{{\mu }{\nu }}$ represents the 6$
\times $6 covariant components of the inertia--metric tensor of
the total mass moving in the knee joint, while $a^{{\nu }}$
corresponds to the 6 contravariant components of the knee
SE(3)--acceleration vector-field.

Now, the covariant (absolute, Bianchi) time--derivative $\frac{{D}}{dt}%
(\cdot )$ of the covariant SE(3)--force $F_{{\mu }}$ defines the
corresponding knee SE(3)--jolt co-vector field:
\begin{equation}
\frac{{D}}{dt}(F_{{\mu }})=m_{{\mu }{\nu }}\frac{{D}}{dt}(a^{{\nu }})=m_{{%
\mu }{\nu }}\left( \dot{a}^{{\nu }}+\Gamma _{\mu \lambda }^{{\nu }}a^{{\mu }%
}a^{{\lambda }}\right) ,  \label{Bianchi}
\end{equation}%
where ${\frac{{D}}{dt}}{(}a^{{\nu }})$ denotes the 6 contravariant
components of the knee SE(3)--jerk vector-field and overdot ($%
\dot{~}$) denotes the time derivative. $\Gamma _{\mu \lambda
}^{{\nu }}$ are the Christoffel's symbols of the Levi--Civita
connection for the SE(3)--group, which are zero in case of pure
Cartesian translations and nonzero in case of rotations as well as
in the full--coupling of translations and rotations.

In the following, we elaborate on the knee SE(3)--jolt concept
(using vector and tensor methods) and its biophysical consequences
in the form of the knee dislocations and disclinations.

\subsection{$SE(3)-$group of local joint motions}

Briefly, the $SE(3)-$group of knee motions is defined as a
semidirect (noncommutative) product of 3D knee rotations and 3D
knee micro--translations,
\begin{equation*}
SE(3):=SO(3)\rhd \mathbb{R}^{3}.
\end{equation*}%
Its most important subgroups are the following:
\begin{center}
{{\frame{$
\begin{array}{cc}
\mathbf{Subgroup} & \mathbf{Definition} \\ \hline
\begin{array}{c}
SO(3),\text{ group of rotations} \\
\text{in 3D (a spherical joint)}%
\end{array}
&
\begin{array}{c}
\text{Set of all proper orthogonal } \\
3\times 3-\text{rotational matrices}%
\end{array}
\\ \hline
\begin{array}{c}
SE(2),\text{ special Euclidean group} \\
\text{in 2D (all planar motions)}%
\end{array}
&
\begin{array}{c}
\text{Set of all }3\times 3-\text{matrices:} \\
\left[
\begin{array}{ccc}
\cos \theta & \sin \theta & r_{x} \\
-\sin \theta & \cos \theta & r_{y} \\
0 & 0 & 1%
\end{array}
\right]%
\end{array}
\\ \hline
\begin{array}{c}
SO(2),\text{ group of rotations in 2D} \\
\text{subgroup of }SE(2)\text{--group} \\
\text{(a revolute joint)}%
\end{array}
&
\begin{array}{c}
\text{Set of all proper orthogonal } \\
2\times 2-\text{rotational matrices} \\
\text{ included in }SE(2)-\text{group}%
\end{array}
\\ \hline
\begin{array}{c}
\mathbb{R}^{3},\text{ group of translations in 3D} \\
\text{(all spatial displacements)}%
\end{array}
& \text{Euclidean 3D vector space}%
\end{array}
$}}}
\end{center}

In other words, the gauge $SE(3)-$group of knee Euclidean
micro-motions contains matrices of the form {\small $\left(
\begin{array}{cc}
\mathbf{R} & \mathbf{p} \\
0 & 1%
\end{array}%
\right) ,$} where $\mathbf{p}$ is knee 3D micro-translation vector
and $\mathbf{R}$ is knee 3D rotation matrix, given by the product $%
\mathbf{R}=R_{\varphi }\cdot R_{\psi }\cdot R_{\theta }$ of the
three
Eulerian knee rotations, $\text{roll}=R_{\varphi },~\text{pitch}%
=R_{\psi },~\text{yaw}=R_{\theta }$, performed respectively about the $x-$%
axis by an angle $\varphi ,$ about the $y-$axis by an angle $\psi ,$ and
about the $z-$axis by an angle $\theta $ (see \cite{9,ParkChung,IJHR}),
{\small
\begin{equation*}
R_{\varphi }=\left[
\begin{array}{ccc}
1 & 0 & 0 \\
0 & \cos \varphi & -\sin \varphi \\
0 & \sin \varphi & \cos \varphi%
\end{array}%
\right] ,~~R_{\psi }=\left[
\begin{array}{ccc}
\cos \psi & 0 & \sin \psi \\
0 & 1 & 0 \\
-\sin \psi & 0 & \cos \psi%
\end{array}%
\right] ,~~R_{\theta }=\left[
\begin{array}{ccc}
\cos \theta & -\sin \theta & 0 \\
\sin \theta & \cos \theta & 0 \\
0 & 0 & 1%
\end{array}%
\right] .
\end{equation*}%
}

Therefore, natural knee $SE(3)-$dynamics is given by the coupling
of Newtonian (translational) and Eulerian (rotational) equations
of the knee motion.

\subsection{Local joint $SE(3)-$dynamics}

To support our locally--coupled loading--rate hypothesis, we formulate the
coupled Newton--Euler dynamics of the knee motions within the $%
SE(3)- $group. The forced Newton--Euler equations read in vector (boldface)
form
\begin{eqnarray}
\text{Newton} &:&~\mathbf{\dot{p}}~\mathbf{\equiv M\dot{v}=F+p\times \omega }%
,  \label{vecForm} \\
\text{Euler} &:&~\mathbf{\dot{\pi}}~\mathbf{\equiv I\dot{\omega}=T+\pi
\times \omega +p\times v},  \notag
\end{eqnarray}%
where $\times $ denotes the vector cross product,
\begin{equation*}
\mathbf{M}\equiv M_{ij}=diag\{m_{1},m_{2},m_{3}\}\qquad \text{and}\qquad
\mathbf{I}\equiv I_{ij}=diag\{I_{1},I_{2},I_{3}\},\qquad (i,j=1,2,3)
\end{equation*}%
are the total moving segment's (diagonal) mass and inertia matrices,\footnote{%
In reality, mass and inertia matrices ($\mathbf{M,I}$) are not
diagonal but rather full $3\times 3$ positive--definite symmetric
matrices with coupled mass-- and inertia--products. Even more
realistic, fully--coupled mass--inertial properties of a moving
segment are defined by the single
non-diagonal $6\times 6$ positive--definite symmetric mass--inertia matrix $%
\mathcal{M}_{SE(3)}$, the so-called material metric tensor of the $SE(3)-$%
group, which has all nonzero mass--inertia coupling products.
However, for simplicity, in this paper we shall consider only the
simple case of two separate diagonal $3\times 3$ matrices
($\mathbf{M,I}$).} defining the total moving segment mass--inertia
distribution, with principal inertia moments given in Cartesian
coordinates ($x,y,z$) by volume integrals
\begin{equation*}
I_{1}=\iiint \rho (z^{2}+y^{2})dxdydz,~~I_{2}=\iiint \rho
(x^{2}+z^{2})dxdydz,~~I_{3}=\iiint \rho (x^{2}+y^{2})dxdydz,
\end{equation*}%
dependent on the knee density $\rho =\rho (x,y,z)$,
\begin{equation*}
\mathbf{v}\equiv v^{i}=[v_{1},v_{2},v_{3}]^{t}\qquad \text{and\qquad }%
\mathbf{\omega }\equiv {\omega }^{i}=[\omega _{1},\omega _{2},\omega
_{3}]^{t}
\end{equation*}%
(where $[~]^{t}$ denotes the vector transpose) are linear
and angular knee--velocity vectors (that
is, column vectors),
\begin{equation*}
\mathbf{F}\equiv F_{i}=[F_{1},F_{2},F_{3}]\qquad \text{and}\qquad \mathbf{T}%
\equiv T_{i}=[T_{1},T_{2},T_{3}]
\end{equation*}%
are gravitational and other external force and torque co-vectors
(that is, row vectors) acting on the knee,
\begin{eqnarray*}
\mathbf{p} &\equiv &p_{i}\equiv \mathbf{Mv}%
=[p_{1},p_{2},p_{3}]=[m_{1}v_{1},m_{2}v_{2},m_{2}v_{2}]\qquad \text{and} \\
\mathbf{\pi } &\equiv &\pi _{i}\equiv \mathbf{I\omega }=[\pi _{1},\pi
_{2},\pi _{3}]=[I_{1}\omega _{1},I_{2}\omega _{2},I_{3}\omega _{3}]
\end{eqnarray*}%
are linear and angular knee--momentum co-vectors.

In tensor form, the forced Newton--Euler equations (\ref{vecForm}) read
\begin{eqnarray*}
\dot{p}_{i} &\equiv &M_{ij}\dot{v}^{j}=F_{i}+\varepsilon _{ik}^{j}p_{j}{%
\omega }^{k},\qquad(i,j,k=1,2,3) \\
\dot{\pi}_{i} &\equiv &I_{ij}\dot{\omega}^{j}=T_{i}+\varepsilon _{ik}^{j}\pi
_{j}\omega ^{k}+\varepsilon _{ik}^{j}p_{j}v^{k},
\end{eqnarray*}
where the permutation symbol $\varepsilon _{ik}^{j}$ is\ defined as
\begin{equation*}
\varepsilon _{ik}^{j}=
\begin{cases}
+1 & \text{if }(i,j,k)\text{ is }(1,2,3),(3,1,2)\text{ or }(2,3,1), \\
-1 & \text{if }(i,j,k)\text{ is }(3,2,1),(1,3,2)\text{ or }(2,1,3), \\
0 & \text{otherwise: }i=j\text{ or }j=k\text{ or }k=i.%
\end{cases}%
\end{equation*}

In scalar form, the forced Newton--Euler equations (\ref{vecForm}) expand as
\begin{eqnarray}
\text{Newton} &:&\left\{
\begin{array}{c}
\dot{p}_{_{1}}={F_{1}}-{m_{3}}{v_{3}}{\omega _{2}}+{m_{2}}{v_{2}}{\omega _{3}%
} \\
\dot{p}_{_{2}}={F_{2}}+{m_{3}}{v_{3}}{\omega _{1}}-{m_{1}}{v_{1}}{\omega _{3}%
} \\
\dot{p}_{_{3}}={F_{3}}-{m_{2}}{v_{2}}{\omega _{1}}+{m_{1}}{v_{1}}{\omega _{2}%
}%
\end{array}%
\right. ,  \label{scalarForm} \\
\text{Euler} &:&\left\{
\begin{array}{c}
\dot{\pi}_{_{1}}={T_{1}}+({m_{2}}-{m_{3}}){v_{2}}{v_{3}}+({I_{2}}-{I_{3}}){%
\omega _{2}}{\omega _{3}} \\
\dot{\pi}_{_{2}}={T_{2}}+({m_{3}}-{m_{1}}){v_{1}}{v_{3}}+({I_{3}}-{I_{1}}){%
\omega _{1}}{\omega _{3}} \\
\dot{\pi}_{_{3}}={T_{3}}+({m_{1}}-{m_{2}}){v_{1}}{v_{2}}+({I_{1}}-{I_{2}}){%
\omega _{1}}{\omega _{2}}%
\end{array}%
\right. ,  \notag
\end{eqnarray}
showing the moving segment's mass and inertia couplings.

Equations (\ref{vecForm})--(\ref{scalarForm}) can be derived from the
translational + rotational kinetic energy of the moving segment\footnote{%
In a fully--coupled Newton--Euler knee dynamics, instead of
equation (\ref{Ek}) we would have moving segment's kinetic energy
defined by the inner product:
\begin{equation*}
E_{k}=\frac{1}{2}\left[{\mathbf{p}}{\mathbf{\pi }}\left\vert \mathcal{M}%
_{SE(3)}\right.{\mathbf{p}}{\mathbf{\pi }}\right] .
\end{equation*}%
}
\begin{equation}
E_{k}={\frac{1}{2}}\mathbf{v}^{t}\mathbf{Mv}+{\frac{1}{2}}\mathbf{\omega }%
^{t}\mathbf{I\omega },  \label{Ek}
\end{equation}%
or, in tensor form
\begin{equation*}
E={\frac{1}{2}}M_{ij}{v}^{i}{v}^{j}+{\frac{1}{2}}I_{ij}{\omega}%
^{i}{\omega}^{j}.
\end{equation*}

For this we use the \emph{Kirchhoff--Lagrangian equations} (see, e.g., \cite%
{Kirchhoff,naomi97}, or the original work of Kirchhoff in German)
\begin{eqnarray}
\frac{d}{{dt}}\partial _{\mathbf{v}}E_{k} &=&\partial _{\mathbf{v}%
}E_{k}\times \mathbf{\omega }+\mathbf{F},  \label{Kirch} \\
{\frac{d}{{dt}}}\partial _{\mathbf{\omega }}E_{k} &=&\partial _{\mathbf{%
\omega }}E_{k}\times \mathbf{\omega }+\partial _{\mathbf{v}}E_{k}\times
\mathbf{v}+\mathbf{T},  \notag
\end{eqnarray}
where $\partial _{\mathbf{v}}E_{k}=\frac{\partial E_{k}}{\partial \mathbf{v}}%
,~\partial _{\mathbf{\omega }}E_{k}=\frac{\partial E_{k}}{\partial \mathbf{%
\omega }}$; in tensor form these equations read
\begin{eqnarray*}
\frac{d}{dt}\partial _{v^{i}}E &=&\varepsilon _{ik}^{j}\left( \partial
_{v^{j}}E\right) \omega ^{k}+F_{i}, \\
\frac{d}{dt}\partial _{{\omega }^{i}}E &=&\varepsilon _{ik}^{j}\left(
\partial _{{\omega }^{j}}E\right) {\omega }^{k}+\varepsilon _{ik}^{j}\left(
\partial _{v^{j}}E\right) v^{k}+T_{i}.
\end{eqnarray*}

Using (\ref{Ek})--(\ref{Kirch}), linear and angular knee--momentum
co-vectors are defined as
\begin{equation*}
\mathbf{p}=\partial _{\mathbf{v}}E_{k}{,\qquad \mathbf{\pi }=\partial _{%
\mathbf{\omega }}E_{k},}
\end{equation*}%
or, in tensor form
\begin{equation*}
p_{i}=\partial _{v^{i}}E{,\qquad }\pi _{i}=\partial _{{\omega }^{i}}E,
\end{equation*}%
with their corresponding time derivatives, in vector form
\begin{equation*}
~\mathbf{\dot{p}}=\frac{d}{dt}\mathbf{p=}\frac{d}{dt}\partial _{\mathbf{v}}E{%
,\qquad \mathbf{\dot{\pi}}=}\frac{d}{dt}\mathbf{\pi =}\frac{d}{dt}\partial _{%
\mathbf{\omega }}E,
\end{equation*}%
or, in tensor form
\begin{equation*}
~\dot{p}_{i}=\frac{d}{dt}p_{i}=\frac{d}{dt}\partial _{v^{i}}E{,\qquad \dot{%
\pi}_{i}=}\frac{d}{dt}\pi _{i}=\frac{d}{dt}\partial _{{\omega }^{i}}E,
\end{equation*}%
or, in scalar form
\begin{equation*}
\mathbf{\dot{p}}=[\dot{p}_{1},\dot{p}_{2},\dot{p}_{3}]=[m_{1}\dot{v}%
_{1},m_{2}\dot{v}_{2},m_{3}\dot{v}_{3}],\qquad {\mathbf{\dot{\pi}}}=[\dot{\pi%
}_{1},\dot{\pi}_{2},\dot{\pi}_{3}]=[I_{1}\dot{\omega}_{1},I_{2}\dot{\omega}%
_{2},I_{3}\dot{\omega}_{3}].
\end{equation*}

While healthy knee $SE(3)-$dynamics is given by the coupled
Newton--Euler micro--dynamics, the knee injury is actually caused
by the sharp and discontinuous change in this natural $SE(3)$
micro-dynamics, in the form of the $SE(3)-$jolt, causing
discontinuous knee deformations, both translational dislocations
and rotational disclinations.

\subsection{Joint injury dynamics: the $SE(3)-$jolt}

The $SE(3)-$jolt, the actual cause of the knee injury (in the form
of the plastic deformations), is defined as a coupled Newton+Euler
jolt; in (co)vector form the $SE(3)-$jolt reads\footnote{%
Note that the derivative of the cross--product of two vectors follows the
standard calculus product--rule: $\frac{d}{dt}(\mathbf{u\times v})=\mathbf{%
\dot{u}\times v+u\times \dot{v}.}$}
\begin{equation*}
SE(3)-\text{jolt}:\left\{
\begin{array}{l}
\text{Newton~jolt}:\mathbf{\dot{F}=\ddot{p}-\dot{p}\times \omega -p\times
\dot{\omega}}~,\qquad \\
\text{Euler~jolt}:\mathbf{\dot{T}=\ddot{\pi}}~\mathbf{-\dot{\pi}\times
\omega -\pi \times \dot{\omega}-\dot{p}\times v-p\times \dot{v}},%
\end{array}
\right.
\end{equation*}
where the linear and angular jolt co-vectors are
\begin{equation*}
\mathbf{\dot{F}\equiv M\ddot{v}}=[\dot{F}_{{1}},\dot{F}_{{2}},\dot{F}_{{3}%
}],\qquad \mathbf{\dot{T}\equiv I\ddot{\omega}}=[\dot{T}_{{1}},\dot{T}_{{2}},%
\dot{T}_{{3}}],
\end{equation*}
where
\begin{equation*}
\mathbf{\ddot{v}}=[\ddot{v}_{{1}},\ddot{v}_{{2}},\ddot{v}_{{3}}]^{t},\qquad
\mathbf{\ddot{\omega}}=[\ddot{\omega}_{{1}},\ddot{\omega}_{{2}},\ddot{\omega}%
_{{3}}]^{t},
\end{equation*}
are linear and angular jerk vectors.

In tensor form, the $SE(3)-$jolt reads\footnote{%
In this paragraph the overdots actually denote the absolute Bianchi
(covariant) time-derivative (\ref{Bianchi}), so that the jolts retain the
proper covector character, which would be lost if ordinary time derivatives
are used. However, for the sake of simplicity and wider readability, we
stick to the same overdot notation.}
\begin{eqnarray*}
~\dot{F}_{i} &=&\ddot{p}_{i}-\varepsilon _{ik}^{j}\dot{p}_{j}{\omega }%
^{k}-\varepsilon _{ik}^{j}p_{j}{\dot{\omega}}^{k}, \qquad(i,j,k=1,2,3) \\
~\dot{T}_{{i}} &=&\ddot{\pi}_{i}~-\varepsilon _{ik}^{j}\dot{\pi}_{j}\omega
^{k}-\varepsilon _{ik}^{j}\pi _{j}{\dot{\omega}}^{k}-\varepsilon _{ik}^{j}%
\dot{p}_{j}v^{k}-\varepsilon _{ik}^{j}p_{j}\dot{v}^{k},
\end{eqnarray*}
in which the linear and angular jolt covectors are defined as
\begin{eqnarray*}
\mathbf{\dot{F}} &\equiv &\dot{F}_{i}=\mathbf{M\ddot{v}}\,\equiv \mathbf{\,}%
M_{ij}\ddot{v}^{j}=[\dot{F}_{1},\dot{F}_{2},\dot{F}_{3}], \\
\mathbf{\dot{T}} &\equiv &\dot{T}_{{i}}=\mathbf{I\ddot{\omega}\equiv \,}%
I_{ij}\ddot{\omega}^{j}=[\dot{T}_{{1}},\dot{T}_{{2}},\dot{T}_{{3}}],
\end{eqnarray*}
where \ $\mathbf{\ddot{v}}=\ddot{v}^{{i}},$ and $\mathbf{\ddot{\omega}}=%
\ddot{\omega}^{{i}}$ are linear and angular jerk vectors.

In scalar form, the $SE(3)-$jolt expands as
\begin{eqnarray*}
\text{Newton~jolt} &:&\left\{
\begin{array}{l}
\dot{F}_{{1}}=\ddot{p}_{1}-m_{{2}}\omega _{{3}}\dot{v}_{{2}}+m_{{3}}\left( {%
\omega }_{{2}}\dot{v}_{{3}}+v_{{3}}\dot{\omega}_{{2}}\right) -m_{{2}}v_{{2}}{%
\dot{\omega}}_{{3}}, \\
\dot{F}_{{2}}=\ddot{p}_{2}+m_{{1}}\omega _{{3}}\dot{v}_{{1}}-m_{{3}}\omega _{%
{1}}\dot{v}_{{3}}-m_{{3}}v_{{3}}\dot{\omega}_{{1}}+m_{{1}}v_{{1}}\dot{\omega}%
_{{3}}, \\
\dot{F}_{{3}}=\ddot{p}_{3}-m_{{1}}\omega _{{2}}\dot{v}_{{1}}+m_{{2}}\omega _{%
{1}}\dot{v}_{{2}}-v_{{2}}\dot{\omega}_{{1}}-m_{{1}}v_{{1}}\dot{\omega}_{{2}},%
\end{array}
\right. \\
&& \\
\text{Euler~jolt} &:&\left\{
\begin{array}{l}
\dot{T}_{{1}}=\ddot{\pi}_{1}-(m_{{2}}-m_{{3}})\left( v_{{3}}\dot{v}_{{2}}+v_{%
{2}}\dot{v}_{{3}}\right) -(I_{{2}}-I_{{3}})\left( \omega _{{3}}\dot{\omega}_{%
{2}}+{\omega }_{{2}}{\dot{\omega}}_{{3}}\right) , \\
\dot{T}_{{2}}=\ddot{\pi}_{2}+(m_{{1}}-m_{{3}})\left( v_{{3}}\dot{v}_{{1}}+v_{%
{1}}\dot{v}_{{3}}\right) +(I_{{1}}-I_{{3}})\left( {\omega }_{{3}}{\dot{\omega%
}}_{{1}}+{\omega }_{{1}}{\dot{\omega}}_{{3}}\right) , \\
\dot{T}_{{3}}=\ddot{\pi}_{3}-(m_{{1}}-m_{{2}})\left( v_{{2}}\dot{v}_{{1}}+v_{%
{1}}\dot{v}_{{2}}\right) -(I_{{1}}-I_{{2}})\left( {\omega }_{{2}}{\dot{\omega%
}}_{{1}}+{\omega }_{{1}}{\dot{\omega}}_{{2}}\right).%
\end{array}
\right.
\end{eqnarray*}

We remark here that the linear and angular momenta ($\mathbf{p,\pi }$),
forces ($\mathbf{F,T}$) and jolts ($\mathbf{\dot{F},\dot{T}}$) are
co-vectors (row vectors), while the linear and angular velocities ($\mathbf{%
v,\omega }$), accelerations ($\mathbf{\dot{v},\dot{\omega}}$) and jerks ($%
\mathbf{\ddot{v},\ddot{\omega}}$) are vectors (column vectors). This
bio-physically means that the `jerk' vector should not be confused with the
`jolt' co-vector. For example, the `jerk'\ means shaking the head's own
mass--inertia matrices (mainly in the atlanto--occipital and atlanto--axial
joints), while the `jolt'means actually hitting the head with some external
mass--inertia matrices included in the `hitting'\ SE(3)--jolt, or hitting
some external static/massive body with the head (e.g., the ground --
gravitational effect, or the wall -- inertial effect). Consequently, the
mass-less `jerk' vector\ represents a (translational+rotational) \textit{%
non-collision effect} that can cause only soft knee injuries,
while the inertial `jolt'\ co-vector represents a
(translational+rotational) \textit{collision effect} that can
cause hard knee injuries.

\subsection{Joint disclinations and dislocations caused by the $SE(3)-$jolt}

For mild knee injury (caused by internal loss of stability), the best injury predictor is considered to
be the product of localized knee strain and strain rate, which is
the standard isotropic viscoelastic continuum concept (see, e.g. \cite{Halloran}). To improve
this standard concept, in this subsection, we consider the knee
joint as a 3D anisotropic multipolar \emph{Cosserat viscoelastic
continuum} \cite{Cosserat1,Cosserat2,Eringen02}, exhibiting
coupled--stress--strain elastic properties. This non-standard
continuum model is suitable for analyzing plastic (irreversible)
deformations and fracture mechanics \cite{Bilby} in multi-layered
materials with microstructure (in which slips and bending of
layers introduces additional
degrees of freedom, non-existent in the standard continuum models; see \cite%
{Mindlin65,Lakes85} for physical characteristics and \cite{Yang81,Yang82},%
\newline
\cite{Park86} for biomechanical applications).

The $SE(3)-$jolt $(\mathbf{\dot{F},\dot{T}})$ causes two types of
localized knee discontinuous deformations:

\begin{enumerate}
\item The Newton jolt $\mathbf{\dot{F}}$ can cause severe micro-translational \emph{%
dislocations}, or discontinuities in the Cosserat translations;

\item The Euler jolt $\mathbf{\dot{T}}$ can cause mild micro-rotational \emph{%
disclinations}, or discontinuities in the Cosserat rotations.
\end{enumerate}

For general treatment on dislocations and disclinations related to
asymmetric discontinuous deformations in multipolar materials, see, e.g.,
\cite{Jian95,Yang01}.

To precisely define the knee dislocations and disclinations,
caused by the $SE(3)-$jolt $(\mathbf{\dot{F},\dot{T}})$, we first
define the coordinate co-frame, i.e., the set of basis 1--forms
$\{dx^{i}\}$, given in local coordinates
$x^{i}=(x^{1},x^{2},x^{3})=(x,y,z)$, attached to the moving
segment's center-of-mass. Then, in the coordinate co-frame
$\{dx^{i}\}$ we introduce the following set of the knee
plastic--deformation--related $SE(3)-$based differential $p-$forms (see \cite%
{14,17}):\newline
$~~~~$the \emph{dislocation current }1--form, $\mathbf{J}=J_{i}\,dx^{i};$%
\newline
$~~~~$the \emph{dislocation density }2--form, $\mathbf{\alpha }=\frac{1}{2}%
\alpha _{ij}\,dx^{i}\wedge dx^{j};$\newline
$~~~~$the \emph{disclination current }2--form, $\mathbf{S}=\frac{1}{2}%
S_{ij}\,dx^{i}\wedge dx^{j};$ ~and\newline
$~~~~$the \emph{disclination density }3--form, $\mathbf{Q}=\frac{1}{3!}%
Q_{ijk}\,dx^{i}\wedge dx^{j}\wedge dx^{k}$,

\noindent where $\wedge $ denotes the exterior wedge--product. According to
Edelen \cite{Edelen,Kadic}, these four $SE(3)-$based differential forms
satisfy the following set of continuity equations:
\begin{eqnarray}
&&\mathbf{\dot{\alpha}}=\mathbf{-dJ-S,}  \label{dis1} \\
&&\mathbf{\dot{Q}}=\mathbf{-dS,}  \label{dis2} \\
&&\mathbf{d\alpha }=\mathbf{Q,}  \label{dis3} \\
&&\mathbf{dQ}=\mathbf{0,}\qquad  \label{dis4}
\end{eqnarray}
where $\mathbf{d}$ denotes the exterior derivative.

In components, the simplest, fourth equation (\ref{dis4}), representing the
\emph{Bianchi identity}, can be rewritten as
\begin{equation*}
\mathbf{dQ}=\partial _{l}Q_{[ijk]}\,dx^{l}\wedge dx^{i}\wedge dx^{j}\wedge
dx^{k}=0,
\end{equation*}
where $\partial _{i}\equiv\partial /\partial x^{i}$, while $\theta _{\lbrack
ij...]}$ denotes the skew-symmetric part of $\theta _{ij...}$.

Similarly, the third equation (\ref{dis3}) in components reads
\begin{eqnarray*}
\frac{1}{3!}Q_{ijk}\,dx^{i}\wedge dx^{j}\wedge dx^{k} &=&\partial _{k}\alpha
_{\lbrack ij]}\,dx^{k}\wedge dx^{i}\wedge dx^{j},\text{\qquad or} \\
Q_{ijk} &=&-6\partial _{k}\alpha _{\lbrack ij]}.
\end{eqnarray*}

The second equation (\ref{dis2}) in components reads
\begin{eqnarray*}
\frac{1}{3!}\dot{Q}_{ijk}\,dx^{i}\wedge dx^{j}\wedge dx^{k} &=&-\partial
_{k}S_{[ij]}\,dx^{k}\wedge dx^{i}\wedge dx^{j},\text{\qquad or} \\
\dot{Q}_{ijk} &=&6\partial _{k}S_{[ij]}.
\end{eqnarray*}

Finally, the first equation (\ref{dis1}) in components reads
\begin{eqnarray*}
\frac{1}{2}\dot{\alpha}_{ij}\,dx^{i}\wedge dx^{j} &=&(\partial _{j}J_{i}-%
\frac{1}{2}S_{ij})\,dx^{i}\wedge dx^{j},\text{\qquad or} \\
\dot{\alpha}_{ij}\, &=&2\partial _{j}J_{i}-S_{ij}\,.
\end{eqnarray*}

In words, we have:

\begin{itemize}
\item The 2--form equation (\ref{dis1}) defines the time derivative $\mathbf{%
\dot{\alpha}=}\frac{1}{2}\dot{\alpha}_{ij}\,dx^{i}\wedge dx^{j}$ of the
dislocation density $\mathbf{\alpha }$ as the (negative) sum of the
disclination current $\mathbf{S}$ and the curl of the dislocation current $%
\mathbf{J}$. This time derivative is caused by the translational part of the SE(3) jolt.

\item The 3--form equation (\ref{dis2}) states that the time derivative $%
\mathbf{\dot{Q}=}\frac{1}{3!}\dot{Q}_{ijk}\,dx^{i}\wedge dx^{j}\wedge dx^{k}$
of the disclination density $\mathbf{Q}$ is the (negative) divergence of the
disclination current $\mathbf{S}$. This time derivative is caused by the rotational part of the SE(3) jolt.

\item The 3--form equation (\ref{dis3}) defines the disclination density $%
\mathbf{Q}$ as the divergence of the dislocation density $\mathbf{\alpha }$,
that is, $\mathbf{Q}$ is the \emph{exact} 3--form.

\item The Bianchi identity (\ref{dis4}) follows from equation (\ref{dis3})
by \textit{Poincar\'{e} lemma} \cite{14,17} and states that the disclination
density $\mathbf{Q}$ is conserved quantity, that is, $\mathbf{Q}$ is the
\emph{closed} 3--form. Also, every 4--form in 3D space is zero.
\end{itemize}

From these equations, we can conclude that the knee dislocations
and disclinations are mutually coupled by the underlaying
$SE(3)-$group, which means that we cannot separately analyze
translational and rotational knee injuries. This result supports the validity of the combined loading hypothesis.

\section{Conclusion}

Based on the previously developed \emph{covariant force law} \cite{12,17}, its recent application to traumatic brain injury \cite{GaneshTBI}, and using as an example a human knee joint, in this
paper we have formulated a new coupled loading--rate hypothesis
for the generic musculo-skeletal injury. This new injury hypothesis states that generic cause of human joint
injuries is an external $SE(3)-$jolt, an impulsive loading hitting
a joint in several degrees-of-freedom, both rotational and
translational, combined and simultaneously. To demonstrate this, we have developed
the vector Newton--Euler mechanics on the Euclidean $SE(3)-$group
of the knee micro-motions. In this way, we have precisely defined
the concept of the $SE(3)-$jolt, which is a cause of two kinds of
rapid joint discontinuous deformations: (i) mild rotational
disclinations, and (ii) severe translational dislocations. Based
on the presented model, we argue that we cannot separately analyze
localized joint rotations from translations, as they are in reality
coupled. To prevent human musculo-skeletal injuries we need to develop the
\textit{musculo-skeletal SE(3)--jolt awareness}, e.g. never overload a flexed knee, avoid any kind of collisions.

\end{document}